\documentstyle[twocolumn,aps]{revtex}

\newcommand{\beq}{\begin{equation}}
\newcommand{\eeq}{\end{equation}}
\newcommand{\beqa}{\begin{eqnarray}}
\newcommand{\eeqa}{\end{eqnarray}}

\newcommand{\abs}[1]{\left| #1 \right|}
\newcommand{\absq}[1]{\abs{#1}^2}

\begin{document}
\draft
\title{Comprehensive experimental test of quantum erasure}
\author{Alexei Trifonov,\thanks{Electronic address: alexei@ele.kth.se \hfil \break Permanent address: Ioffe Physical Technical Institute, 26 Polytekhnicheskaya, 194021 St. Petersburg, Russia.} Gunnar Bj\"{o}rk, Jonas S\"{o}derholm, and Tedros Tsegaye}
\address{Department of Electronics, Royal Institute of Technology (KTH), Electrum 229, SE-164 40 Kista, Sweden}
\date{\today}
\maketitle

\begin{abstract}
In an interferometer, path information and interference
visibility are incompatible quantities. Complete determination of
the path will exclude any possibility of interference, rendering
the visibility zero. However, if the composite object and probe
state is pure, it is, under certain conditions, possible to trade
the path information for improved (conditioned) visibility. Such
a procedure is called quantum erasure. We have performed such
experiments with polarization entangled photon pairs. Using a
partial polarizer we could vary the degree of entanglement between
object and probe. We could also vary the interferometer splitting
ratio and thereby vary the {\em a priori} path predictability.
We have tested quantum erasure under a number of different
experimental conditions and found good agreement between
experiments and theory.
\end{abstract}


\narrowtext

\section{Introduction}

A fundamental difference between classical physics and quantum
mechanics is that the latter, being a linear dynamical theory,
allows superpositions. The superposition principle, in turn, leads
directly to the concept of complementarity, the fact that any
quantum system has at least two properties that cannot
simultaneously be known. Complementarity
\cite {Heisenberg,Bohr} has been discussed intensively since
the early development of quantum mechanics. Recently some new
qualitative statements about complementarity have been proposed
\cite
{Wootters,Greenberger,Mandel,Walther,Raymer,Bhandari,Tan,Jaeger,Englert,Bjork},
and subsequently the question has been raised whether there
exist any relations between these new expressions and the
Schr\"{o}dinger-Robertson and the Arthurs-Kelley uncertainty
principles \cite {Bhandari,Tan,Englert,Bjork 2,Storey}. In
addition, the fundamental physical mechanism that enforces
complementarity has been debated \cite {Storey,Englert
2,Wiseman,Luis 3}. The superposition principle also leads to
nonlocality. This consequence of quantum theory led Einstein,
Podolsky and Rosen to write their famous EPR-paper \cite{Einstein}, arguing that
this consequence of quantum theory rendered the theory
``unreasonable''. Later Bell devised an
experimental procedure in which predictions of quantum mechanics
could be tested against predictions based on locally realistic
theories \cite{Bell}. Several such experiments have been made
\cite {Aspect,Kiess,Tapster,Fry,Weihs} and all results have been
compatible with the predictions of quantum mechanics.

When deriving his uncertainty principle, Heisenberg erroneously
attributed the uncertainty to the back-action on the measured
object from the measurement apparatus. Later work has clarified
that Heisenberg's uncertainty relation only makes a statement
about the preparation of a quantum mechanical state. If one wants
to qualitatively record the back-action on the state due to a
measurement of some observable $\hat{A}$, one also has to measure
the conjugate observable to $\hat{A}$ on the ``same'' state.
(Unless the state is an eigenstate of $\hat{A}$ the state will
change as a result of the $\hat{A}$-measurement, therefore we
have put the word ``same'' in quotes.) This is
often called a simultaneous measurement, i.e., on every state in an
ensemble two incompatible measurements are made. To qualitatively
describe the uncertainty product of the two simultaneous
measurements one arrives at quantitatively or even qualitatively
different uncertainty principles from that of Heisenberg
\cite{Bjork 2,Arthurs,Arthurs 2,She,Stenholm,Martens,Appleby}.
What is surprising is that the
back-action, or at least the measured uncertainty due to the back-action, is
not solely a property of the object, and the object and probe
interaction, but depends also on how the probe is measured and
how the obtained information is used. Under certain
circumstances, the apparent indeterminism of the object due to
the measurement back-action can be undone by the action of a
{\em local} operation on the probe. This procedure is called
quantum erasing \cite{Scully} and is a manifestation of the
nonlocality of quantum mechanics. Various implementations of such
experiments, and their connection to complementarity have been
discussed in some recent papers \cite{Walther,Bjork,Bjork
2,Kwiat2}. Several experiments have also been performed \cite
{Kwiat1,Herzog,Pittman,Rempe,Kim,Ou,Zou}. What distinguishes our
experiment from the previous ones is that we have been able to
vary both the degree of object and probe entanglement and,
independently, the {\em a priori} path information. This leads to
a more complex situation than previously reported.

In order to discuss quantum erasure in greater detail we will start
by making a few definitions. Quantum mechanics only makes
predictions about states. It does not say anything about paths,
modes or objects. All these words are classical concepts, but are
nonetheless useful in discussions about quantum erasure.
Usually quantum erasure is discussed in the context of an object
taking one of two paths in an interferometer (or passing through
one of two slits in a Young's double-slit experiment). In a more
strict manner of speaking the two paths are described by two
orthogonal modes. The object, and its path is then defined by two
state vectors $| O_+ \rangle$ and $| O_-
\rangle$. In general this description is insufficient to describe
all possible outcomes of an experiment. In this case it is more
proper to define the object and its path by two sets of states
$\{|O_{+,i}\rangle\}$ and $\{|O_{-,i}\rangle\}$, where all states
in one set are mutually orthonormal (and quite obviously all
pairs of states from different sets are orthogonal). This is, e.g.,
the case in a recent experiment by Schwindt {\em et al.}
\cite{Schwindt}, where the paths were defined in terms of two
spatial modes, and in each of the paths the object could be
either in a vertically or in a horizontally polarized state (or in
a superposition or mixture of the two polarization states). However, to
describe our experiment we shall only need to consider a
two-dimensional object Hilbert space
${\cal H}_O $. Note that in defining the object in terms of a
pair of orthonormal states, the concept of ``path'' should not be
taken literally. In our experiment the two modes corresponding to
the ``paths'' are actually two orthogonal linear polarization
modes, or equivalently, two orthogonal polarization states in the
same spatial mode.

To determine the state of the object, and hence identify the
``path'', we need only a two dimensional probe Hilbert space
${\cal H}_M$, spanned by the two orthonormal state vectors $| M_+
\rangle$ and $| M_-
\rangle$. In order to make any statement about the object's ``path''
through a measurement of the probe, the object and the probe
need to be in an entangled state described by the density matrix
$\hat{\rho}$ belonging to the space ${\cal H}_O \otimes
{\cal H}_M$. In our experiment the preparation of $\hat{\rho}$ is
accomplished by a combination of photon-pair generation in a
frequency down-converting nonlinear crystal and a partial
polarizer interacting with the object.

The only {\em a priori} information we have are known the probabilities $w_+$ and $w_- = 1-w_+$
for the two events. (These are given by the object state
preparation.) The Maximum Likelihood (ML) estimation strategy
(which is one of many possible strategies) dictates that we
should, for each and every event, guess that the object took the
most likely ``path''. The strategy will maximize the likelihood
$L$ of guessing correctly. The likelihood will be
$L = {\rm Max} \{ w_+, w_-\}$, and from this relation it is
evident that $1/2 \leq L \leq 1$. The likelihood can be
renormalized to yield the predictability $P$
\cite{Jaeger,Englert}, given by
\begin{equation}
P = 2 L -1 .
\label{eq:disting}
\end{equation}
It is clear that $0 \leq P \leq 1$, where $P=0$ corresponds to a random
guess of which ``path'' the object took, and $P=1$ corresponds to
absolute certainty about the ``path''.

One can also compute the visibility when the two ``path'' probability
amplitudes interfere. The visibility $V$, too, is a statistical measure
which requires an ensemble of identically prepared systems to estimate. The
classical definition of $V$ is
\begin{equation}
V = {\frac{I_{\rm max}-I_{\rm min} }{I_{\rm max}+I_{\rm min}}} ,
\label{eq:classical visibility definition}
\end{equation}
where $I_{\rm max}$ and $I_{\rm min}$ are the intensities of the interference fringe
maxima and minima. For a single quanta we can only talk about the
probability $p$ of the object being detected on a specific
location on a screen, or exiting one of two interferometer ports.
(Do not confuse this probability
$p$ with the predictability $P$.) The probability $p$ will vary essentially
sinusoidally with position on the screen, or with the interferometer
arm-length difference. In this case the natural definition of $V$ is
\begin{equation}
V = {\frac{p_{\rm max}-p_{\rm min} }{p_{\rm max}+p_{\rm min}}} .
\label{eq:quantum visibility definition}
\end{equation}

It has been shown \cite{Englert} that $P$ and $V$ for an object
whose ``path'' is determined by one of two orthonormal states,
satisfies the following inequality:
\begin{equation}
P^{2}+V^{2}\leq 1,
\label{eq:Complementarity relation}
\end{equation}
where the upper bound is saturated for any pure state. Note that if one wants to verify the unequality (\ref{eq:Complementarity relation}) experimentally, one needs two ensembles of identical states. On the first ensemble one makes a ``path" measurement and on the second one makes a visibility measurement. Hence, on any one state only one (sharp) measurement is performed.

\section{Probing the ``path''}

In order to retrieve more information about the ``path'' of the
object then what is given by {\em a priori} knowledge of
 ``path'' probabilities (\ref{eq:disting}), it is possible to
use an ancillary probe system. By using a correlated ancillary
system one can simultaneously get ``path'' information (from the
probe ancilla) and visibility information (from the object). In
the case where the Hilbert space of the object is two-dimensional,
the probe must possess at least two degrees of freedom. Hence, in
the simplest case the total system belongs to a 2 $\times$ 2 dimensional
composite Hilbert space. The measurement of ``path'' information
and/or visibility must be preceded by an interaction between the
object and probe which entangles their degrees of freedom. Hence
the simplest state after the interaction will be a four-mode
state consisting of the object, whose ``path'' and visibility
information of we wish to measure, and the probe, which will help
us to get information about the object's ``path''. We assume that an
interaction between the object and probe leaves the probabilities
$w_{+}$ and $w_{-}$ invariant. This is not the most general
entangling interaction possible, it defines the subset of
entangling interactions of the quantum non-demolition (QND) kind
\cite{Bjork}.

A few different experimental situations can be distinguished:

1. The state after the interaction can be factorized in the two composite
Hilbert spaces ${\cal H}_{O}$ and ${\cal H}_{M}$. Then both
systems can be treated independently. The state of the probe
carries no information about the object and vice versa. This is a
trivial and not particularly interesting situation.

2. The state is a perfectly entangled state so that the probe
contains full ``path'' information of the object. Thus no
interference between the object ``path'' probability amplitudes can
take place. Our ability to predict which ``path'' the object took is
perfect, while the visibility is zero. It is possible, however,
to retrieve the object preinteraction ``path'' interference
visibility by doing a conditioned measurement. To retrieve the
visibility, it is necessary to give up the ``path'' information.
Therefore the probe must be measured in such a way that
the information encoded in the state of the probe is not revealed
by the measurement, i.e., the probe must be measured in such a way
that the corresponding observable is complementary to
the observable that discloses the ``path''. For a pure state it is
thus possible to restore the visibility of the initial state
(conditioned visibility) with subsequent loss of the ``path''
information.

3. The state is partially entangled. This is an intermediate case between
the previous two. Only partial information about the ``path'' of the
object can
be extracted from the probe. That still leaves room for non-zero
 ``path'' visibility. This intermediate case is examined carefully in
the paper.

\bigskip The most general scheme of the measuring procedure is shown in
Fig. 1. The object can take one of two ``paths'' and the probe is
used to determine which ``path'' the object took. Before the
interaction (plane A) the object and the probe are independent
and the state is represented by a product of the corresponding
density operators:

\begin{equation}
\hat{\rho}=\hat{\rho}_{O} \otimes \hat{\rho}_{M} ,
\label{in_state}
\end{equation}
where $\hat{\rho}_{O}=\left| \Psi _{O}\right\rangle \left\langle
\Psi _{O}\right| $ and $\left| \Psi _{O}\right\rangle
=\sqrt{w_{+}}\left| O_{+}\right\rangle +e^{i\phi
}\sqrt{w_{-}}\left| O_{-}\right\rangle $. This is not the most
general density operator since it represents a pure state, but
since only pure states saturate Eq. (\ref{eq:Complementarity
relation}) we will concentrate on this case in this work. The
role of the interaction is to entangle the object and the probe.
We assume that the interaction affects only the probe's degrees
of freedom and works like a unitary transformation

\begin{eqnarray}
\hat{U}\left| O_{+}\right\rangle \left| M\right\rangle &=&\left|
O_{+}\right\rangle \left| m_{+}\right\rangle ,
\label{unitary_transf} \\
\hat{U}\left| O_{-}\right\rangle \left| M\right\rangle &=&\left|
O_{-}\right\rangle \left| m_{-}\right\rangle , \nonumber
\end{eqnarray}
where $\left| M\right\rangle$ is the initial probe
state. The state after the interaction (plane B) becomes

\begin{equation}
\left| \Psi _{e}\right\rangle =\sqrt{w_{+}}\left|
O_{+}\right\rangle \left| m_{+}\right\rangle +e^{i\phi
}\sqrt{w_{-}}\left|O_{-}\right\rangle \left|
m_{-}\right\rangle .
\label{ent_state}
\end{equation}
The corresponding density operator is denoted $\hat{\rho}_e$.
As mentioned above, this is not the most general case of entanglement (the general
case is discussed in \cite{Bjork}) but sufficient for our task.
If $\left\langle m_{+}|m_{-}\right\rangle =0$ (perfect
entanglement) then the ``path'' the object took can be extracted
perfectly from a measurement of the probe. It is convenient to
introduce
\begin{equation}
c=\left|
\left\langle m_{+}|m_{-}\right\rangle
\right|
\label{eq:c from rho}
\end{equation}
as a measure of entanglement. Note that $c$
depends both on the state $\hat{\rho}_{e}$ and also, in general,
on the choice of the object basis. However, since the object
basis is fixed, (the ``path'' is defined in terms of the fixed
states
$|O_{+}\rangle$ and $|O_{-}\rangle$), $c$ is an unambiguous
measure of the entanglement. If there is no entanglement then
$\left| m_{+}\right\rangle =\left| m_{-}\right\rangle$, which
implies $c=1$.

From an experimental point of view it is more convenient to deal
with an orthogonal probe basis $\left| M_{+}\right\rangle $ and
$\left| M_{-}\right\rangle $. It is always possible to choose, for
simplicity, $\left\langle m_{+}|M_{-}\right\rangle =0$. In this
new basis

\begin{eqnarray}
\left| \Psi _{e}\right\rangle & = & \sqrt{w_{+}}\left| O_{+}\right\rangle \left|
M_{+}\right\rangle +e^{i\phi }c\sqrt{w_{-}}\left| O_{-}\right\rangle \left|
M_{+}\right\rangle \nonumber \\
& & + e^{i\phi }\sqrt{w_{-} (1-c^{2})} \left|
O_{-}\right\rangle \left| M_{-}\right\rangle .
\label{partial_ent_state}
\end{eqnarray}

One of the simplest experimental realizations of this state
is a superposition of two
single-photon two-mode states. Unfortunately the strength of the
state-of-the-art nonlinear interaction at single photon level is
too weak in order to produce the state (\ref{partial_ent_state})
from (\ref{in_state}) by interaction of object and probe
photons. It is tempting to try however to simulate the state
(\ref{partial_ent_state}) as a result of spontaneous parametric
down conversion (SPDC). In the process of SPDC the pump photon is
split into a pair of photons. In Sec. 4 we will show
that under a specific condition the state
(\ref{partial_ent_state}) can be produced. But first let us
introduce the quantities of interest for a quantitative
discussion of ``path'' and visibility information.

\section{Distinguishability and visibility }

The complementary nature of the object before the interaction is
fully described by $P$ and $V$. The predictability can be
computed from $\hat{\rho}_e$ as:
\begin{eqnarray}
P & \equiv & |w_+ - w_-| \nonumber \\
& = & \left| \left\langle O_{+}\right|
{\rm Tr}_{M}\{\hat{\rho}_e\}\left| O_{+}\right\rangle -\left\langle
O_{-}\right| {\rm Tr}_{M}\{\hat{\rho}_e\}\left| O_{-}\right\rangle
\right| ,
\label{eq:P from rho}
\end{eqnarray}
where the trace is taken over the probe Hilbert space. By our
choice of interaction (QND-type of entanglement) the
predictability remains invariant, but the post-interaction
visibility will, in general be smaller than the preinteraction
visibility. The visibility can also be computed from $\hat{\rho}_e$
\begin{equation}
V=2\left| \left\langle O_{+}\right| {\rm
Tr}_{M}\{\hat{\rho}_e\}\left| O_{-}\right\rangle \right| .
\label{eq:V from rho}
\end{equation}

These expressions are consistent with Eqs. (\ref{eq:disting}) and
(\ref{eq:quantum visibility definition}). Since the visibility
is not a conserved quantity, we will denote the preinteraction
visibility $V_{0}$. In general
$V\leq V_{0}$, which can be attributed to random relative-phase
shifts associated with the interaction process \cite{Luis 3}.

The measure of the post-interaction ``path'' information is the
distinguishability which is given by

\begin{equation}
D={\rm Tr}_{M}\left \{ \left\| \left\langle O_{+}\right|
\hat{\rho}_e\left| O_{+}\right\rangle -\left\langle O_{-}\right|
\hat{\rho}_e\left| O_{-}\right\rangle \right\| \right \},
\label{eq:D from rho}
\end{equation}
where $\left\| a\right\| $ denotes the trace-class norm of
$\hat{a}$ (see for example \cite{Klauder}). Choosing the
entanglement interaction in the way we did ensures that $P\leq
D$. As it was shown by Englert \cite{Englert}, complementarity
leads to the inequality

\begin{equation}
D^{2}+V^{2}\leq 1 .
\label{eq:Englert distinguishability equation}
\end{equation}
This expression has a clear physical explanation: It is possible to get more
information about the ``path'' ($D$) only on the expense of the
conjugate observable, which is the relative-phase and is
quantified by ($V$) for the discussed case \cite{Bjork 2,Luis 3}. It
means that $D$ contains both the {\em a priori} ``path''
information as well as the additional information encoded in the
state of the probe. It should be noted that the
distinguishability denotes the maximum information about the
``path'' that can be extracted from the probe by a measuring
apparatus. Obtaining the full information can be accomplished for
example by optimal probe state projection on photodetectors by
adjusting the unitary evolution \
$\hat{U}_{M}$ preceding the photodetectors (see Fig. 1). For an
arbitrary, in general non-optimal probe measurement basis, the
quantitative measure of obtained ``path'' information is the
so-called measured distinguishability \cite{Bjork} (the same
quantity is called ``knowledge" by Englert \cite{Schwindt}). It
can mathematically be expressed:

\begin{eqnarray}
\lefteqn{D_{m} = \left| \left\langle M_{+}\right| \hat{U}_m \left( \left\langle O_{+}\right| \hat{\rho}_e \left| O_{+}\right\rangle -\left\langle O_{-}\right| \hat{\rho}_e \left| O_{-}\right\rangle \right) \hat{U}_m^\dagger \left| M_{+}\right\rangle \right|} & & \nonumber \\
& & +\left| \left\langle M_{-}\right| \hat{U}_m \left( \left\langle O_{+}\right| \hat{\rho}_e \left| O_{+}\right\rangle -\left\langle O_{-}\right| \hat{\rho}_e \left| O_{-}\right\rangle \right) \hat{U}_m^\dagger \left| M_{-}\right\rangle \right| . \label{eqn: Dm}
\end{eqnarray}

The associated visibility, conditioned on the outcome of the probe measurement,
can similarly be expressed:
\begin{eqnarray}
V_c & = & 2 \left| \left\langle M_{+}\right| \hat{U}_m \left\langle O_{+}\right|
 \hat{\rho}_e \left| O_{-}\right\rangle \hat{U}_m^\dagger \left|
M_{+}\right\rangle \right| \nonumber \\
& & + 2 \left| \left\langle M_{-}\right| \hat{U}_m \left\langle O_{+}\right|
 \hat{\rho}_e \left| O_{-}\right\rangle \hat{U}_m^\dagger \left|
M_{-}\right\rangle \right| .
\label{eqn:Vc}
\end{eqnarray}

To interpret the equation above in terms of a concrete measurement procedure,
the visibility data should be sorted in two sets depending on the outcome of 
the probe measurement. The conditioned visibility then is the probability 
weighted average of the two obtained visibilities. It can be seen directly from 
the form of (\ref{eq:V from rho}) and (\ref{eqn:Vc}) that $V \leq V_c \leq V_0$. The relations between 
all quantities for a pure state are summarized in Table I and by the inequalities (\ref{eq:Complementarity relation})
and (\ref{eq:Englert distinguishability equation}).

As it was shown in \cite{Bjork 2} there exists
a mutual symmetry between $P$ and $V_{0}$. The symmetry implies
that that the visibility can be treated as predictability if the
observables corresponding to the ``path'' measurement and the
visibility measurement are exchanged. However, entanglement
of the kind (\ref{unitary_transf}) breaks this symmetry: with the chosen
entanglement it is feasible to get more information about the
 ``path'' $D\geq P$ (see Table I) but at the expense of the
visibility. In the ``best'' case (optimal quantum erasure) it is
possible only to restore initial visibility. This follows from
our choice of entanglement. We optimized the interaction to use
the probe to encode the ``path'' information and not the
complementary observable, which is quantified by the visibility.

\section{Experimental test of complementarity}

A pair of photons entangled in polarization was used to produce
the initial state (\ref{partial_ent_state}). The experimental
setup is shown in Fig. 2. It is similar to that described in
\cite{Kwiat3}. The main difference is the use of a pulsed pump
source (a Ti:Sapphire laser emitting at 780 nm) incorporating frequency doubling to 390 nm. The length of the pump
pulse was chosen to be 1 ps which is long enough to minimize the
group velocity dispersion problem. The use of a pulsed source
reduced substantially the random coincidence counts between
signal photons and dark counts and between coincident dark
counts. Thus no dark-count corrections were made in any of the
data presented in this paper. A beta-barium borate (BBO) crystal
with type I phase matching was used for frequency doubling and
a type II BBO crystal was used for SPDC. Special care was taken to
compensate for the BBO group velocity mismatch. Photodetection was
accomplished by EG\&G single photon detectors with around 60\% quantum
efficiency. Identical 10 nm bandpass filters
were placed in front of each of the detectors to select
degenerate photon pairs. Two polarizers were used to select
linearly polarized photons. Polarization rotation was
accomplished by two $\lambda /2$ plates placed before the
polarizers. A 94\% fringe visibility in coincidence measurement
was observed when one $\lambda /2$-plate was fixed at 22.5 deg
with respect to the BBO principal axes (horizontal-vertical)
and the other $\lambda /2$-plate was rotated.

\subsection{A maximally entangled state}

Let us start with the state, which was produced in the
experiment by the BBO-crystal (followed by state postselection to
eliminate the predominant $\left| 0,0\right\rangle $-state):

\begin{equation}
\left| \Psi ^{-}\right\rangle =\frac{1}{\sqrt{2}}\left ( \left|
\uparrow ,\rightarrow \right\rangle -\left| \rightarrow
,\uparrow
\right\rangle \right ) ,
\end{equation}
or in a more suitable notation

\begin{equation}
\left| \Psi ^{-}\right\rangle =\frac{1}{\sqrt{2}}\left ( \left|
O_{+},M_{+}\right\rangle -\left| O_{-},M_{-}\right\rangle \right
) .
\label{eq:rotated state}
\end{equation}
This state corresponds to the second of the experimental
situations listed in Sec. 2. For such a state $P=0$, $D=1$, and
$V=0$. The measured distinguishability $D_{m}\left( \theta
\right) $ and the conditioned visibility $V_{c}\left( \theta
\right) $ were measured with respect to the probe basis
rotation angle $\theta $:

\begin{eqnarray}
\left( \matrix{\left| M_{+}(\theta )\right. \rangle \cr \left| M_{-}(\theta )\right. \rangle} \right) & = & \hat{U}^\dagger_m \left( \matrix{\left| M_{+}\right. \rangle \cr \left| M_{-}\right. \rangle} \right) \nonumber \\
& = & \left( \matrix{\cos \theta & \sin \theta \cr -\sin \theta & \cos \theta} \right) \left( \matrix{\left| M_{+}\right. \rangle \cr \left| M_{-}\right. \rangle} \right) ,
\end{eqnarray}
where the angle $\theta$ refers to the rotation angle from the
horizontal plane. If the rotated probe measurement basis is used, the state (\ref{eq:rotated state}) is written:

\begin{eqnarray}
\left| \Psi ^{-}\right\rangle & = & \frac{1}{\sqrt{2}} \{ \cos \theta \left[
\left| O_{+},M_{+}\left( \theta \right) \right\rangle -\left|
O_{-},M_{-}\left( \theta \right) \right\rangle \right] \nonumber \\
& & -\sin \theta \left[
\left| O_{+},M_{-}\left( \theta \right) \right\rangle -\left|
O_{-},M_{+}\left( \theta \right) \right\rangle \right] \} .
\end{eqnarray}
After a calculation using Eqs. (\ref{eqn: Dm}) and (\ref{eqn:Vc}) one finds $D_{m}\left(
\theta \right) =\left| \cos \theta \right| $ and $V_{c}\left(
\theta \right) =\left| \sin \theta \right| $. Results of the
measurements are shown in Fig. 3. Remember that the visibility of this state is zero for any probe measurement basis rotation. The degradation of the conditioned visibility due to the
non-perfect mode overlap was taken into account in plotting
the solid lines. This was done by multiplying the theoretically predicted conditioned visibility $V_c$ by 0.94 which was the maximum experimentally obtained visibility. (For the particular entanglement we chose, the
distinguishability relies on energy and momentum conservation,
whereas good visibility also requires good mode overlap. The
aforementioned group velocity dispersion associated with short
orthogonally polarized photon pulses prevented us from getting
perfect visibility as can be seen in the figure. Hence, the measured $D_M$
goes from very close to unity to zero, whereas $V_c$ goes from zero but only reaches 0.94 at its maximum point.) The apparent error in the figure is larger than the measurement error of $D_m$ and $V_c$ since it is the squares of these quantities, rather than the quantities themselves, that are plotted. We beg the reader to keep this point in mind in the following.

\subsection{A partially entangled state}

A partial polarizer was inserted in the object beam rotated
at an angle $\alpha$ with respect to the horizontal plane (see
Fig. 4). The partial polarizer consisted of a stack of $N$ glass
plates held at the Brewster angle with respect to the stack
rotation axis. The amplitude transmittivities of the partial polarizer were
$t_{\rm p}\approx 1$ and $t_{\rm s}=t$, where $t$ was determined by the
number of plates $N$. (The indices p and s refer to the
linear polarization states with the respect of the partial
polarizer.)

In order to calculate the state after the polarizer it is
convenient to rotate both bases so that the $\left|
O_{+}\right\rangle $\ state
becomes parallel to the p-plane of the partial polarizer

\begin{equation}
\left| \Psi ^{-}\right\rangle =\frac{1}{\sqrt{2}}\left ( \left| O_{+}\left(
\alpha \right) ,M_{+}\left( \alpha \right) \right\rangle -\left| O_{-}\left(
\alpha \right) ,M_{-}\left( \alpha \right) \right\rangle \right )
.
\end{equation}
Now it is easy to find the state after the partial polarizer:
\begin{equation}
\left| \Psi _{e}\right\rangle =\frac{1}{\sqrt{1+t^{2}}}\left ( \left|
O_{+}\left( \alpha \right) ,M_{+}\left( \alpha \right) \right\rangle
-t\left| O_{-}\left( \alpha \right) ,M_{-}\left( \alpha \right)
\right\rangle \right ) .
\end{equation}
A backward rotation by $-\alpha$ gives

\begin{eqnarray}
\left| \Psi _{e}\right\rangle & = & a_{1}\left| O_{+},M_{+}\right\rangle
-a_{2}\left| O_{-},M_{-}\right\rangle \nonumber \\
& & + a_{3}\left( \left|
O_{+},M_{-}\right\rangle - \left| O_{-},M_{+}\right\rangle \right) ,
\label{eq:state from t and alpha}
\end{eqnarray}
where
\begin{eqnarray}
a_{1} = \frac{t+\left( 1-t\right) \cos ^{2}\alpha
}{\sqrt{1+t^{2}}} , \\
a_{2} = \frac{t+\left( 1-t\right) \sin ^{2}\alpha
}{\sqrt{1+t^{2}}} , 
\end{eqnarray}
and
\begin{equation}
a_{3} = \frac{\left( 1-t\right) \sin \alpha \cos \alpha }{\sqrt{1+t^{2}}} .
\end{equation}

Now let us find the correlation coefficients with respect to the probe basis
rotation. As $\theta$ is the angle of the probe measurement basis
rotation, the state can be written as

\begin{eqnarray}
\left| \Psi _{e}\right\rangle & = & b_{1}\left| O_{+},M_{+}\right\rangle
-b_{2}\left| O_{-},M_{-}\right\rangle +b_{3}\left| O_{+},M_{-}\right\rangle \nonumber \\
& & + b_{4}\left| O_{-},M_{+}\right\rangle ,
\label{eq:final entangled state}
\end{eqnarray}
where $b_{1}=a_{1}\cos \theta -a_{3}\sin \theta $,
$b_{2}=a_{2}\cos \theta +a_{3}\sin \theta $, $b_{3}=a_{3}\cos
\theta + a_{1}\sin \theta $, and
$b_{4}=a_{3}\sin \theta +a_{2}\cos \theta $. The photocount
coincidence probabilities become

\begin{eqnarray}
P_{++} & = & \left| \left\langle O_{+,}M_{+}\left( \theta \right) |\Psi
_{e}\right\rangle \right| ^{2}= |b_{1}(\theta)|^{2} , \\
P_{--} & = & \left| \left\langle O_{-,}M_{-}\left( \theta \right) |\Psi
_{e}\right\rangle \right| ^{2}= |b_{2}(\theta)|^{2} , \\
P_{+-} & = & \left| \left\langle O_{+,}M_{-}\left( \theta \right)
|\Psi _{e}\right\rangle \right| ^{2}= |b_{3}(\theta)|^{2} ,
\end{eqnarray}
and
\begin{eqnarray}
P_{-+} & = & \left| \left\langle O_{-,}M_{+}\left( \theta \right)
|\Psi _{e}\right\rangle \right| ^{2}= |b_{4}(\theta)|^{2} . 
\end{eqnarray}
If we denote $\theta_0$ to be the probe measurement angle defined
by $b_{3}(\theta_0)=0$, (for a pure state such an angle always
exists) the relation between (\ref{partial_ent_state}) and
(\ref{eq:final entangled state}) is given as

\begin{equation}
w_{+}= |b_1(\theta)|^2 + |b_3(\theta)|^2 = \left| b_{1}\left(
\theta _{0}\right)
\right| ^{2} ,
\label{eq:experimental wplus}
\end{equation}
and
\begin{equation}
c=\sqrt{\frac{\left| b_{4}\left( \theta _{0}\right) \right| ^{2}}{\left|
b_{2}\left( \theta _{0}\right) \right| ^{2}+\left| b_{4}\left( \theta
_{0}\right) \right| ^{2}}} = \sqrt{{|b_4(\theta_0)|^2\over
1-|b_1(\theta_0)|^2}} .
\label{eq:experimental c}
\end{equation}
Knowing the correlation probabilities it is possible to derive quantities associated with the ``path":

\begin{eqnarray}
P & = & \left| w_{+}-w_{-}\right| \nonumber \\
& = & \left| |b_{1}(\theta)|^{2}+|b_{4}(\theta)|^{2}-|b_{2}(\theta)|^{2}
-|b_{3}(\theta)|^{2}\right| ,
\label{eq:P from b} \\
D_{m}(\theta) & = & \left|
|b_{1}(\theta)|^{2}- |b_{3}(\theta)|^{2}\right| +\left|
|b_{4}(\theta)|^{2}- |b_{2}(\theta)|^{2}\right| .
\label{eq:Dm from b}
\end{eqnarray}
Conditioned visibility can be measured in the same way if we change the object detector basis from 0/90 deg to 45/135 deg. In the new basis the coincidence probabilities are
\begin{eqnarray}
P_{45++} & = & \frac{\absq{b_1 (\theta) + b_4 (\theta)}}{2} , \\
P_{45--} & = & \frac{\absq{b_2 (\theta) + b_3 (\theta)}}{2} , \\
P_{45+-} & = & \frac{\absq{b_3 (\theta) - b_2 (\theta)}}{2} , \\
P_{45-+} & = & \frac{\absq{b_4 (\theta) - b_1 (\theta)}}{2} ,
\end{eqnarray}
which makes it possible to calculate the quantities associated with relative phase measurements
\begin{eqnarray}
V & = & \abs{P_{45++} (\theta) - P_{45-+} (\theta) + P_{45+-} (\theta) - P_{45--} (\theta)} , \label{eq:V from Ps} \\
V_c (\theta) & = & \abs{P_{45++} (\theta) - P_{45-+} (\theta)} + \abs{P_{45+-} (\theta) - P_{45--} (\theta)} . \label{eq:Vc from Ps}
\end{eqnarray}

We measured the distinguishability and the conditioned visibility
using eight different combinations of partial polarizer rotation
angles and number of plates. Here we present only two of the
combinations, but all the measurements were in good
agreement with the theory.

\subsubsection{Low {\em a priori} ``path'' information}

The measured and calculated data for a home made, 10 plate,
partial polarizer rotated by 43 degrees with respect to the
horizontal plane are shown in Fig. 5. The 10 plate partial
polarizer corresponds to
$t=0.200$. Using Eqs. (\ref{eq:state from t and alpha}), (\ref{eq:c
from rho}), (\ref{eq:P from rho}), (\ref{eq:V from rho}), and
(\ref{eq:D from rho}), we can compute the relevant parameters of
the state to be $c=0.716$, $P=0.065$, $V=0.925$, and $D=0.381$,
while a direct measurement gives $P=0.070$,
$V=0.940$, and $D=0.367$, via Eqs. (\ref{eq:P from b}), (\ref{eq:Dm from b}), and (\ref{eq:V from Ps}).
($D$ is the maximum of $D_m (\theta)$ while $V$ is the minimum of $V_c (\theta)$.)
The state differs from the previous one in that although the
predictability is almost zero, the distinguishability is nowhere
near unity. That is, the object and the probe are only weakly
entangled. The results for the
$D_{m}$ and $V_{c}$ measurements are shown in Fig. 6. One should
note specifically that as predicted
$D_{m}$ is bound from below by $P$, and from above by
$D$. In the same manner we see that $V_{c}$ is bound from below
by $V$, and from above by $(1-P^2)^{1/2}$.

It should be noted that the primary data, Fig. 5, agrees
better with theory than the secondary data, Fig. 6. We suspect that
the origin of this effect can be traced to our
home made partial polarizer. The partial polarizer glass plates were not mounted perfectly in parallel. In an independent measurement we recorded the transmission of the polarized
laser light (before the frequency doubling) through the partial polarizer as 
a function of the rotation angle $\alpha$. The transmission
curves should be a displaced cosine curve
which was not quite the case for our polarizer. However, since we did not possess
a goniometer it was hard to confirm that the suspected imperfection was indeed the cause, so we have opted to publish the uncorrected data.

We can also note that when the state is not perfectly entangled,
$D_{m}^{2}+V_{c}^{2}$ is not a conserved quantity although the
state is pure, confirming the predictions in \cite{Bjork}. The reason is
that the observable corresponding to $V_c$ in this case
is not strictly complementary to the observable corresponding
to $D_m$. Therefore the distinguishability measurement and
the visibility measurement does not strictly probe complementary
information about the state. However, the sum
$D_{m}^{2}+V_{c}^{2}$ saturates the bound (\ref{eq:Englert distinguishability equation}) when $D_{m}=D$,
in accordance with the predictions of \cite{Bjork}. The
sharp ``corners" in the theoretical curves signify those
rotation angles where the maximum likelihood strategy dictates a
change in how the probe measurement outcomes should be used for the ``path''
estimation.

\subsubsection{High {\em a priori} ``path'' information}

The next example of a complementarity measurement is shown in
Fig. 7 and Fig. 8. In this measurement a seven plate glass stack
was used as a partial polarizer and the angle of rotation was
adjusted to 21 deg. This corresponds to amplitude
transmittivity $t=0.324$. From these data we can calculate that
$c=0.828$, $P=0.643$, $V=0.563$, and $D=0.839$, which are close
 to the values from the experimental values
$P=0.639$, $V=0.550$, and $D=0.839$. This state is characterized
by its large predictability, in contrast to the previously
treated states. Since $D$ must be larger than or equal to $P$, it
also means that the distinguishability is high.

In Fig. 8 an even stranger shape of the function
$D_{m}^{2}+V_{c}^{2}$ is seen. Since the predictability is large,
there is little information to be had from the probe. We see that
for most probe measurement bases the measured distinguishability
cannot be improved beyond the {\em a priori} predictability. Only
within a small interval of probe measurement rotations will the
information encoded in the state of the probe improve the
measured distinguishability, rendering it (at best) equal to
$D$. Similarly, for most rotations the conditioned visibility
does not exceed the visibility $V$. For the meter basis rotations
between about 10 and 30 degrees (in theory) or between 10 and 20
degrees (in the experiment), both the measured distinguishability and
the conditioned visibility attain their respective minimum values
simultaneously. Here, the state is prepared in such a way that
the ``proper'' complementary observable (as defined in
\cite{Bjork 2}) is complementary both to the ``path'' observable
and the conditioned visibility observable. This is a
manifestation of the fact that any Hilbert space of dimension 2
allows three mutually complementary observables.

\section{Conclusions}

The complementarity relation quantitatively derived by Englert
\cite{Englert} between ``path'' information and ``path''
interference visibility was tested under a wide range of
experimental situations. Using a partial polarizer we were able
to generate states with different {\em a priori} ``path''
information and different degrees of entanglement. The experiment
was designed to, as close as possible, be an
implementation of the theory in
\cite{Englert} and \cite{Bjork}. This is in contrast to a recent
experiment by Schwindt {\em et al.} \cite{Schwindt}, where complementarity is tested
without employing entanglement and with a larger object Hilbert
space than the two dimensional space prescribed by the theory in
\cite{Englert}. The latter experiment can be analyzed and fully
understood in terms of classical physics, whereas the experiment
above, and, e.g., the experiments reported in
\cite{Herzog,Pittman,Rempe,Kim} employ entangled states and
hence quantum non-locality. By changing the measurement basis
of the probe (a local operation) and by using conditioned
measurements, the experiments gave us the possibility to verify
the complementarity relations encompassing quantum erasure
\cite{Bjork}. For non-ideal ($P \geq 0$, $D \leq 1$) but pure
composite states, such a test yields rather surprisingly, only
piecewise differentiable curves, reflecting the nonlinear maximum
likelihood estimation strategy underlying the theory. The
experimental results were in good agreement with theoretical
predictions.

\acknowledgments

This work was supported by grants from the Swedish Technical
Science Research Council, the Swedish Natural Science Research
Council, the Royal Swedish Academy of Sciences, and by INTAS
through Grant 167/96.

\section{Captions}

FIG. 1. Schematic setup for a QND-type measurement of the complementary ``path'' and visibility observables. $\hat{U}_m$ symbolizes a local probe unitary transformation before probe state is irrevokably collapsed by the ``which-path'' meter.

FIG. 2. Experimental setup for testing complementarity by the means of photon polarization measurements on maximally entangled photon pairs. The labels SPC, P and $\lambda/2$ signify single photon counters, polarizers and half-wave plates, respectively.

FIG. 3. Results for measured distiguishability and conditioned visibility versus probe meter basis rotation. Lines represent a theoretical simulation.

FIG. 4. Experimental setup for non-perfect QND-type measurements of photon polarization.

FIG. 5. Coincidence probabilities versus probe meter basis rotation after a 10 plate partial polarizer, rotated by $\alpha = 43$ degrees with respect to the horizontal plane, was inserted in the object beam.

FIG. 6. Results for measured distiguishability and conditioned visibility versus meter basis rotation for the case of low {\em a priori} ``path" information.

FIG. 7. Coincidence probabilities versus meter basis rotation after a 7 plate partial polarizer, rotated by $\alpha = 21$ degrees with respect to the horizontal plane, was inserted in the object beam.

FIG. 8. Results for measured distiguishability and conditioned visibility versus probe meter basis rotation for the case of high {\em a priori} ``path" information.

\mediumtext

\begin{table}
\caption{Which-path and visibility quantities, and their mutual relations.}
\begin{tabular}{l c c c c c}
& Quantities determined & & Quantities determined by & & Quantities determined \\
& by the reduced object & & the composite state $\hat{\rho}$ and & & by the composite \\
& state $\hat{\rho}_O = {\rm Tr}_M \{ \hat{\rho} \}$ & & the choice of probe basis & & state $\hat{\rho}$ \\
\tableline
``Which path" & $P$ & $\leq$ & $D_m$ & $\leq$ & D \\
Visibility & $V = \sqrt{1-D^2}$ & $\leq$ & $V_c$ & $\leq$ & $V_0 = \sqrt{1-P^2}$ \\
\end{tabular}
\end{table}

\narrowtext

\end{document}